\documentclass[journal=aaembp, manuscript=letter]{achemso}
\usepackage{chemformula}
\usepackage[utf8]{inputenc}
\usepackage{achemso}

\title{Terahertz rectennas on flexible substrates based on one-dimensional metal-insulator-graphene diodes}

\author{Andreas Hemmetter}
\affiliation{Advanced Microelectronic Center Aachen (AMICA), AMO GmbH, Otto-Blumenthal-Str. 25, 52074 Aachen, Germany}
\alsoaffiliation{Chair of Electronic Devices, Faculty of Electrical Engineering and Information Technology, RWTH Aachen University, Otto-Blumenthal-Str. 2, 52074 Aachen, Germany}

\author{Xinxin Yang}
\affiliation{Department of Microtechnology and Nanoscience, Chalmers University of Technology, SE-41296 Gothenburg, Sweden}

\author{Zhenxing Wang}
\affiliation{Advanced Microelectronic Center Aachen (AMICA), AMO GmbH, Otto-Blumenthal-Str. 25, 52074 Aachen, Germany}
\email{wang@amo.de}

\author{Martin Otto}
\affiliation{Advanced Microelectronic Center Aachen (AMICA), AMO GmbH, Otto-Blumenthal-Str. 25, 52074 Aachen, Germany}

\author{Burkay Uzlu}
\affiliation{Advanced Microelectronic Center Aachen (AMICA), AMO GmbH, Otto-Blumenthal-Str. 25, 52074 Aachen, Germany}
\alsoaffiliation{Chair of Electronic Devices, Faculty of Electrical Engineering and Information Technology, RWTH Aachen University, Otto-Blumenthal-Str. 2, 52074 Aachen, Germany}

\author{Marcel Andree}
\affiliation{Institute for High-Frequency and Communication Technology, University of Wuppertal, 42119 Wuppertal, Germany}

\author{Ullrich Pfeiffer}
\affiliation{Institute for High-Frequency and Communication Technology, University of Wuppertal, 42119 Wuppertal, Germany}

\author{Andrei Vorobiev}
\affiliation{Department of Microtechnology and Nanoscience, Chalmers University of Technology, SE-41296 Gothenburg, Sweden}

\author{Jan Stake}
\affiliation{Department of Microtechnology and Nanoscience, Chalmers University of Technology, SE-41296 Gothenburg, Sweden}

\author{Max C. Lemme}
\affiliation{Advanced Microelectronic Center Aachen (AMICA), AMO GmbH, Otto-Blumenthal-Str. 25, 52074 Aachen, Germany}
\alsoaffiliation{Chair of Electronic Devices, Faculty of Electrical Engineering and Information Technology, RWTH Aachen University, Otto-Blumenthal-Str. 2, 52074 Aachen, Germany}

\author{Daniel Neumaier}
\affiliation{Chair of Smart Sensor Systems, University of Wuppertal, 42119 Wuppertal, Germany}
\alsoaffiliation{Advanced Microelectronic Center Aachen (AMICA), AMO GmbH, Otto-Blumenthal-Str. 25, 52074 Aachen, Germany}

\usepackage{microtype}
\usepackage{mathpazo}

\usepackage{booktabs}
\usepackage{array}
\usepackage{amsmath}
\usepackage[nolists]{endfloat}
\usepackage{siunitx}
\usepackage{amssymb}
\usepackage{tabularx}
\usepackage{caption}
\usepackage{colortbl}
\usepackage[colorlinks]{hyperref}
\mciteErrorOnUnknownfalse

\newcommand{\ohm}{$\Omega$}

\newcolumntype{P}[1]{>{\raggedright}p{#1}}

\begin{document}

\begin{abstract}
Flexible energy harvesting devices fabricated in scalable thin-film processes are important components in the field of wearable electronics and the Internet of Things.
We present a flexible rectenna based on a one-dimensional junction metal-insulator-graphene diode, which offers low-noise power detection at terahertz (THz) frequencies. 
The rectennas are fabricated on a flexible polyimide film in a scalable process by photolithography using graphene grown by chemical vapor deposition.
A one-dimensional junction area reduces the junction capacitance and enables operation in the D-band (110 - 170~GHz).
The rectenna on polyimide shows a maximum voltage responsivity of 80~V/W at 167~GHz in free space measurements and minimum noise equivalent power of 80~pW/$\sqrt{\text{Hz}}$.
\end{abstract}

\section*{}
%Introduction

\begin{figure*}[t]
\includegraphics[width=\linewidth]{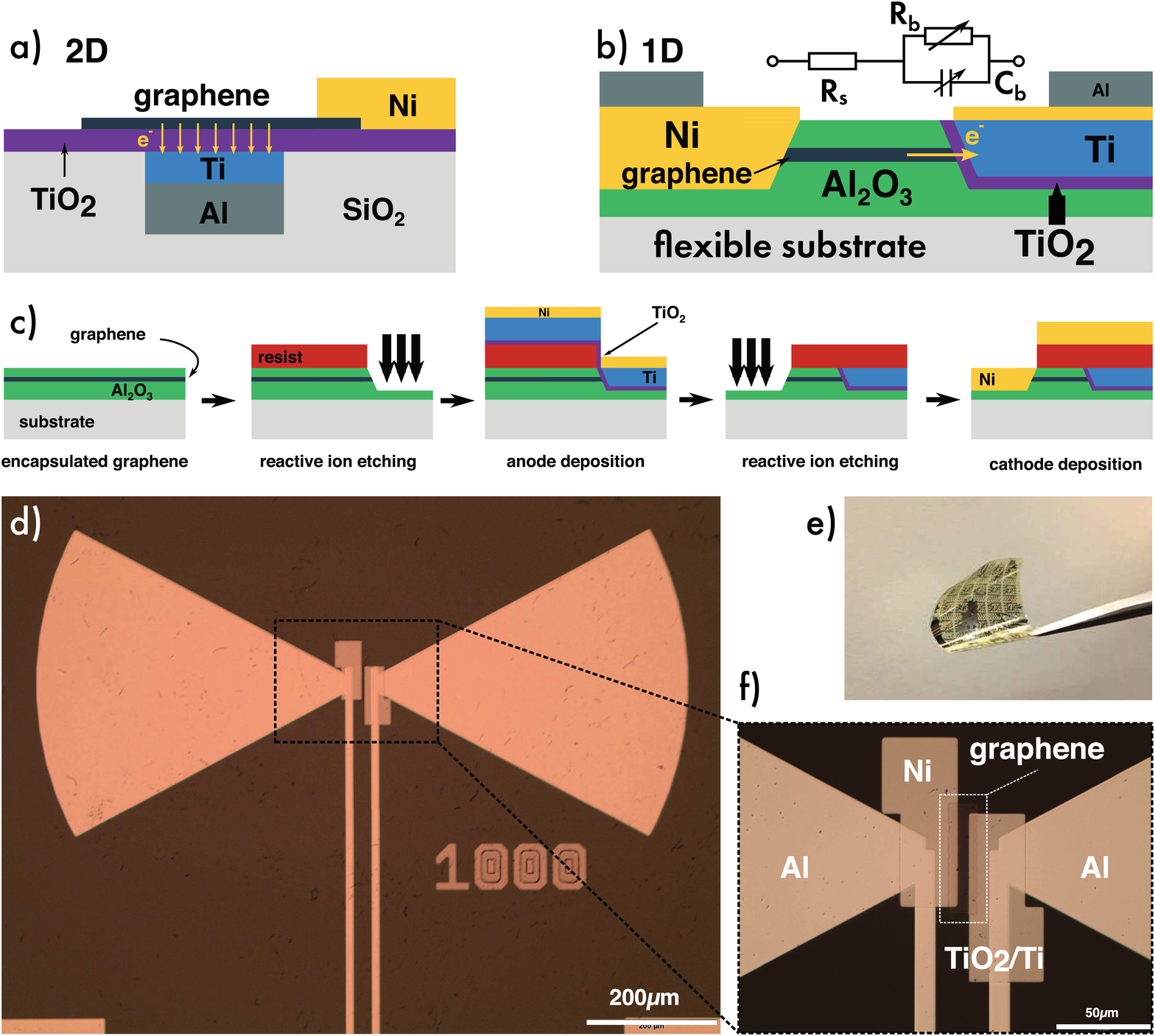}
\caption{Rectenna structure and fabrication. a) Cross-section of a 2D MIG diode compared to a b) 1D MIG diode with equivalent circuit. c) Key fabrication steps for the 1D MIG diode. d) Micrograph of a rectenna on polyimide. e) Photograph of the sample on polyimide after peeling it off the Si carrier substrate. f) Close-up of a 1D MIG diode at the antenna feedpoint.}
\label{fig:sample}
\end{figure*}

Terahertz (THz) radiation is a region of the electromagnetic spectrum with frequencies between 0.1 and 10~THz \cite{tonouchi2007,siegel2002,mittleman2003}.
Miniaturized THz sources and detectors enable a variety of applications, such as communications, surveillance screening, material analysis, biomedical diagnostics and personal healthcare tracking \cite{tonouchi2007, siegel2002, sizov2010, huang2015}.
The emergence of wearable electronics and networks of small, independent sensors for Internet of Things (IoT) applications is driving research both in low power electronic circuits and in energy harvesting on the device or chip level. 
Miniaturized THz power detectors may become crucial components that can function as energy harvesting devices, in particular on flexible thin-film substrates, where they can overcome the form factor limitations of silicon (Si) electronic chips and can be fabricated in scalable roll-to-roll processes. 
Thus, they have the potential to power decentralized sensor networks, passive readout circuits or integrated mobile devices without the need for batteries or an external power supply \cite{bolivar}.\\

\textit{Rectennas} - antenna-coupled diodes - are versatile two-terminal devices that directly rectify a detected signal. 
Their zero-bias operation makes them suitable for energy harvesting applications \cite{moddel2013, zhu2013}. 
This principle has been well established in the microwave region since 1966 \cite{brown66}, due to the availability of Schottky diodes with sufficiently short response times \cite{pfeiffer08, Sankaran20052, Gammon2012, Georgiadou2020}.\\

Increased cutoff frequencies in the THz range \cite{gadalla2014} have been achieved by metal-insulator-metal (MIM) diodes, which utilize tunneling and thermionic majority carrier conduction.
However, MIM diodes generally show inferior DC performance compared to conventional p-n junction or Schottky diodes \cite{Shriwastava2019, Periasamy2011}.
Metal-insulator-\textit{graphene} (MIG) diodes (Figure \ref{fig:sample}a), where the cathode
metal of a MIM diode is replaced by graphene, combine excellent DC performance with high cutoff frequencies \cite{Shriwastava2019,Urcuyo2016,shaygan17,wang19}.
The high charge carrier mobility and flexibility of graphene \cite{wang19} allow such devices to be used in flexible THz rectennas, an application space that has been inaccessible due to the rigidity, bias and fabrication requirements of conventional semiconductor-based detectors.
We present rectennas based on edge-contacted MIG diodes (Figure \ref{fig:sample}b), fabricated in a scalable thin-film compatible process, enabling high-throughput fabrication.
The results point towards the possibility to integrate rectenna arrays as power supplies in flexible, wearable and conformal devices, for example wearable biomedical or distributed environmental sensors.\\

\section*{}
%Theory

The operation frequency of a diode can be estimated using:
\begin{equation}
f_c = \frac{1}{2 \pi R_s C_b}
\label{eq:cutoff}
\end{equation}
where $R_s$ is the access resistance and $C_b$ is the barrier capacitance from the diode's equivalent circuit seen in Figure \ref{fig:sample}b \cite{wang19, Cowley1966}.
The low access resistance of conventional MIM diodes allows them to reach very high cutoff frequencies in the THz range, however with rather poor rectification performance \cite{Shriwastava2019}.
MIG diodes, which have a similar layout as MIM diodes (Figure \ref{fig:sample}a), offer better rectification, but at the expense of a larger access resistance and therefore lower operation speed, as one metal layer is replaced by graphene.
Furthermore, the current through a two-dimensional (2D) MIG diode is emitted perpendicularly to the graphene plane across a small "van-der-Waals gap" and is therefore affected by surface species at the graphene-insulator interface. 
This can increase the junction resistance, hindering an efficient coupling between antenna and diode\cite{shaygan17}.\\

By forming a one-dimensional (1D) diode junction only to the graphene edge, as shown in the schematic cross section in Figure \ref{fig:sample}b, the junction capacitance can be significantly reduced, enabling operation frequencies up to the THz range \cite{wang13}.
The effective junction area is then given by $A = w \cdot t$, where $w$ is the channel width and $t$ is the thickness of graphene, about 0.3~nm \cite{ni2007}.
This leads simultaneously to a reduced junction capacitance  ($C_b$), improved current injection at the insulator-graphene and graphene-metal interfaces and reduced operating voltage due to the electric field enhancement at the 1D edge \cite{wang13}.\\

\section*{}
%Device Fabrication and Design

The rectennas consist of edge-contacted metal-insulator-graphene diodes at the feedpoint of broadband bowtie antennas.
The rectennas have been fabricated on THz-transparent polyimide (PI) films on silicon handling substrates using standard microprocessing techniques such as contact photolithography and reactive ion etching.
Key fabrication steps are shown in Figure \ref{fig:sample}c.
The PI has first been covered with an aluminum oxide (\ch{Al2O3}) layer through plasma-enhanced atomic layer deposition (ALD).
CVD-grown graphene has been transferred in a wet process \cite{delarosa2013} onto the substrate and encapsulated with 20~nm \ch{Al2O3} in a thermal ALD process.
A photolithographically defined mask has been used for reactive ion etching (RIE) through the \ch{Al2O3} and the graphene. The resist remains on the chip at this point and acts as a lift-off mask for the subsequent atomic layer deposition of 5~nm \ch{TiO2} insulator, 20~nm sputtered titanium (Ti) anode metal \cite{shaygan17, wang19} and 13~nm nickel (Ni). The self-aligned nature of the process prevents contamination of the graphene edge from photoresist residues or misalignment of graphene and metal layers, which would increase the diodes' capacitance and resistance. This process results in a one-dimensional MIG junction, because only the graphene edge is in electrical contact with the \ch{TiO2} barrier (Figure \ref{fig:sample}b), as opposed to the two-dimensional contact in conventional MIG stacks (Figure \ref{fig:sample}a).
A 25~nm thick Ni contact to the graphene has been fabricated in a similar way, i.e. by reactive ion etching through the \ch{Al2O3}/graphene/\ch{Al2O3} stack and subsequent metal deposition. 
Here, the graphene edge is in direct contact with the Ni and forms an ohmic contact \cite{wang19, robinson2011, kretz2018}.
At this point, graphene still covers the entire chip apart from directly below both metal contacts.
The remaining graphene has been patterned into 6~$\mu m$ long and 50~$\mu m$ wide channels, before 100~nm thick aluminum (Al) has been deposited as antennas, transmission lines and contact pads.\\

A key feature of the presented metal-insulator-graphene rectenna is that it is compatible with conventional thin-film technologies, i.e. it does not involve high temperature processing steps $>$ 300$^\circ$C and uses only $\mu$m-scale lithography. 
A micrograph of a fabricated rectenna is shown in Figure \ref{fig:sample}d. The polyimide film is removed from the Si carrier substrate after the device fabrication for electrical measurements. Figure \ref{fig:sample}e shows a flexible PI substrate with several THz rectennas. A close-up of the finished diode on a flexible substrate is shown in Figure \ref{fig:sample}f.\\

The transmission behavior of the antennas was designed for broadband absorption to match the measurement setup tailored to the D band from 110 to 170~GHz.
The parameter space was modeled using the electromagnetic finite element solver HFSS with a lumped port ($Z_a$ = 377~\ohm) at the feedpoint.
The chosen bowtie antenna design with an opening angle of 30$^{\circ}$ has a large bandwidth of 43\% at 157~GHz (defined as a voltage standing wave ratio VSWR $\leq$ 2), enabling free space measurements over the D band from 110 to 170~GHz.
A pair of 1~mm long metal traces lead from the diode contacts to larger contact pads for biasing and signal readout.
The transmission line serves as a low-pass filter that rejects the received THz signal and only allows the rectified DC component to pass.\\

\section*{}
%Results and Discussion

\subsection*{}
%DC Characterization

\begin{figure*}[t]
\includegraphics[width=\linewidth]{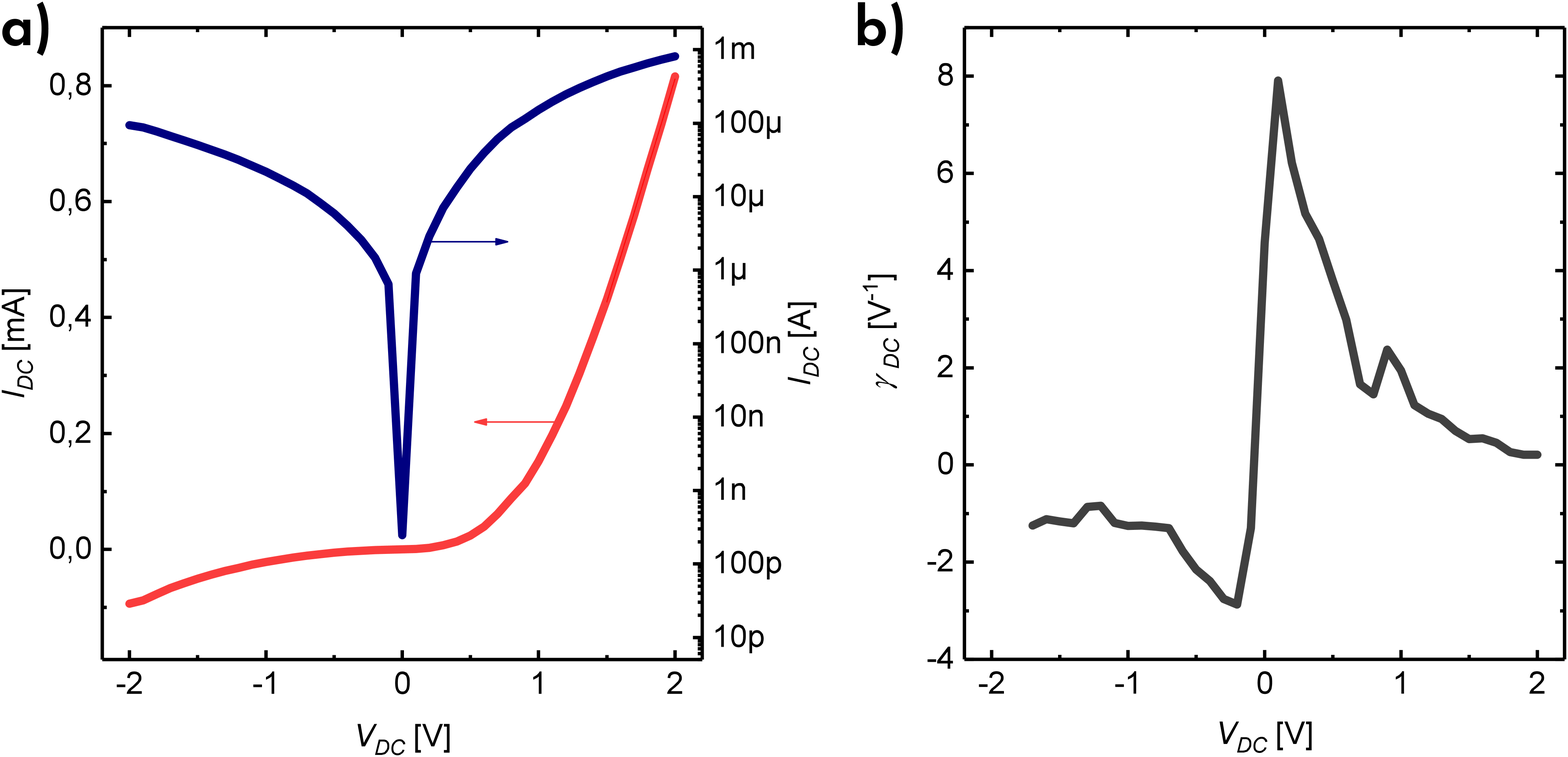}
\caption{DC characteristics of a diode on a polyimide film. a) Linear (blue) and logarithmic (red) $I$-$V$ curve and b) responsivity $\gamma_{DC}$.}
\label{fig:flex-dc}
\end{figure*}

Direct current (DC) characteristics of the rectennas have been measured under ambient conditions.
Current-voltage ($I$-$V$) measurements, where the bias voltage $V_{DC}$ has been swept from -2 to +2~V, are shown linearly and logarithmically in Figure \ref{fig:flex-dc}a. The device shows a non-linear $I$-$V$ characteristic and reaches maximum currents of $I_{DC} = 0.8$~mA at a forward bias voltage of $V_{DC} = 2$~V and $I_{DC} = 0.1$~mA at $V_{DC} = -2$~V. 
Typically, such currents are normalized by the device dimensions in order to compare them to the state of the art. 
Here, only the graphene edge emits charge carriers across the barrier. 
Taking into account the graphene thickness of approximately 0.3~nm and the device width of 50~µm, the resulting maximum current density at the graphene edge exceeds $5 \times 10^6$ A/cm$^2$ in the forward direction, consistent with previously reported 1D-MIG diodes by Wang et al. \cite{wang19}.\\

The $I$-$V$ curve allows calculating the responsivity $\gamma_{DC}$ of the diode, one of the decisive figures of merit for RF applications. 
The responsivity is a measure for the rectification efficiency of the diode \cite{shaygan17}: 
\begin{equation}
    \gamma_{dc} = \frac{\text{d}^2 I/\text{d}V^2}{\text{d}I/\text{d}V}
\end{equation}
As can be seen from its expanded units ($\frac{1}{V} = \frac{A}{V A} = \frac{A}{W}$), a diode's responsivity is related to the rectified current at a given input power.
The flexible diode reaches a maximum responsivity of 8~V$^{-1}$ (Figure \ref{fig:flex-dc}b) in the vicinity of $V_{DC} = 0$~V.\\

\subsection*{}
%THz Measurements

\begin{figure*}[!ht]
\centering \includegraphics[width=\textwidth]{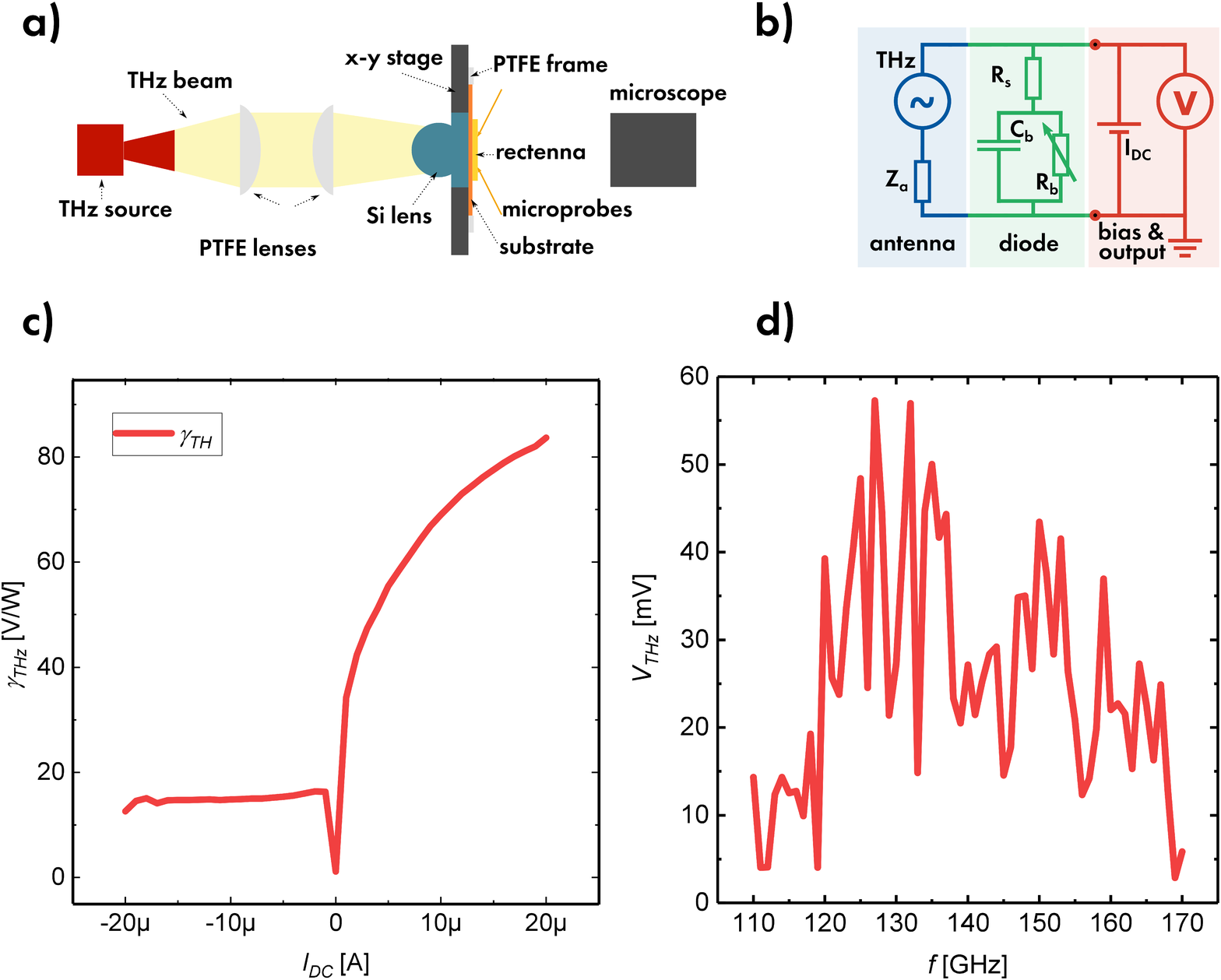}
\caption{THz measurement of a flexible rectenna. a) Schematic of the free-space measurement setup and b) equivalent circuit of the system. c) Bias current dependency of the voltage responsivity. d) Frequency dependency of the rectified voltage under illumination with THz radiation.}
\label{fig:flex-thz}
\end{figure*}

Measurements at THz frequencies between 110 and 170~GHz were conducted in free space under ambient conditions.
A sine-wave modulated THz signal (333~Hz) was transmitted by a horn antenna, collimated and prefocused by two identical PTFE (Teflon) lenses and finally focused onto the back of the sample by a hyper-hemispherical Si lens. 
The setup is shown schematically in Figure \ref{fig:flex-thz}a.
After peeling off the PI film from the Si substrate, the flexible chip was mounted onto a high-resistivity Si lens in the beam path.
Microprobes contacted the sample behind the beam path. 
A Keithley 2604B source meter provided the DC input bias as a current through an external resistor and simultaneously measured the DC voltage at the diode.
The THz source was composed of a VDI WR6.5GX extender driven by an Agilent 83650A signal generator with 30~dB sine wave modulation at 333~Hz.
The rectified voltage response was measured using an SR830 lock-in amplifier. 
For calibration, the available output power of the extender $P_{ava}$ across the D band was measured using an Erickson calorimetric power meter directly at the waveguide flange of the extender on which the horn antenna was mounted.
To describe the diode's behavior during THz measurement we formulate a small-signal equivalent circuit, shown in Figure \ref{fig:flex-thz}b.\\

A key THz detector characteristic is the optical voltage responsivity $\gamma_{THz}$, which is defined in Eq. \ref{eq:rv} as the DC output voltage per THz power and is given in units of [V/W]:
\begin{equation}
\gamma_{THz} = \frac{2 \sqrt{2} \Delta V}{P_{ava}} = \beta_{THz} R
\label{eq:rv}
\end{equation}
Here, $\Delta V$ is the rectified DC voltage measured by the lock-in amplifier \cite{yang17}. 
The optical current responsivity $\beta_{THz}$ in [A/W] can be calculated from the voltage responsivity and the total diode resistance $R$ at the respective operating point.
The pre-factors originate in the peak-to-peak and root mean square (rms) amplitudes of the lock-in amplifier \cite{yang17}. The bias-dependent THz response was measured at 167~GHz with a THz output power $P_{ava} = 1~\text{mW}$. 
The maximum optical voltage responsivity $\gamma_{THz} = 83.7~\text{V/W}$ was reached at an applied bias current of 20~µA and a measured DC voltage of 1.47~V (Figure \ref{fig:flex-thz}c). 
This corresponds to an optical current responsivity $\beta_{THz} = 1.13~\text{mA/W}$ at a total device resistance $R = 74~\text{k\ohm}$.\\

It is noticeable that the optical responsivity under THz measurements increases with applied bias, while the DC responsivity shows a peak close to zero bias.
This can be explained by the non-ideal impedance matching between diode and antenna.
As a higher bias is applied, the junction resistance of the diode decreases, thus better matching its impedance to that of the antenna, and increasing the power delivered to the diode.\\

The rectified voltage over the D band from 110 to 170~GHz is shown in Figure \ref{fig:flex-thz}d.
During the frequency sweep the bias current was fixed at -10~µA and the rectified voltage was recorded.
The voltage $V_{THz}$ does not show any noticeable frequency dependency, but fluctuations, possibly due to interferences in the polyimide film.
The Teflon and silicon lenses used to focus the THz beam contribute an absorption loss of around 3~dB, and a further unknown part of the beam is reflected or scattered at various interfaces without contributing to the incident power.
Even though the substrate materials are transparent to radiation in the THz regime, the mismatch in refractive indices introduces undesirable frequency dependent reflections that reduce the RF intensity at the rectenna.
As such, the measured value is a lower bound of the optical responsivity.\\

\begin{figure*}[!ht]
\centering \includegraphics[width=\textwidth]{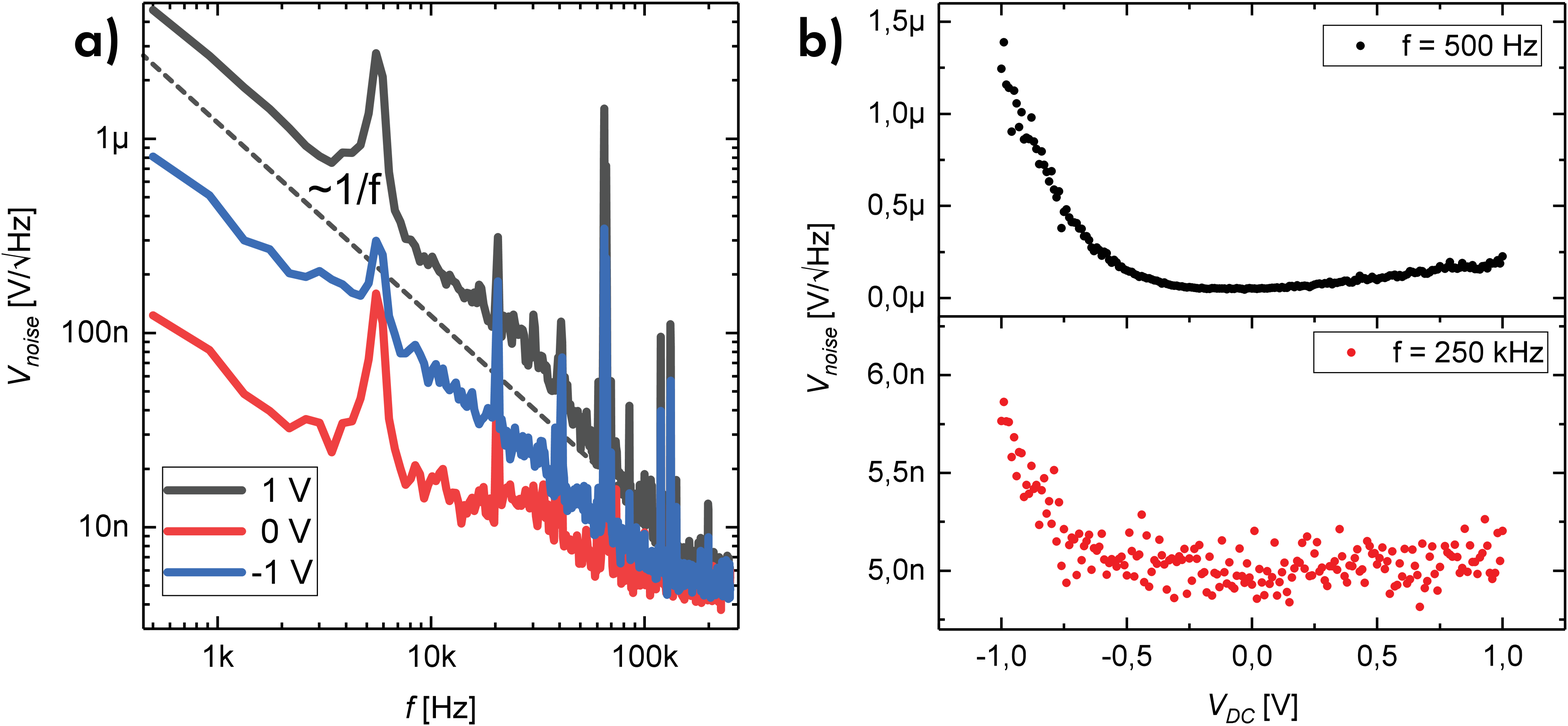}
\caption{Noise measurements. a) Voltage noise spectrum from 500~Hz to 250~kHz, measured on-chip. b) Bias dependence of the voltage noise at 500~Hz (top) and 250~kHz (bottom).}
\label{fig:noise-thz}
\end{figure*}

A test device with ground-signal-ground pads and an identical diode to the one in the rectenna has been fabricated to measure the voltage noise $V_{noise}$ directly on-chip. Figure \ref{fig:noise-thz}a shows the noise at different bias levels at frequencies between 500~Hz and 250~kHz. The bias dependence of the voltage noise is shown in Figure \ref{fig:noise-thz}b. The noise equivalent power ($\text{NEP} = {V_{noise}}/{\gamma_{THz}}$) of the detector can be calculated from the voltage noise $V_{noise}$ and the responsivity $\gamma_{THz}$. At 250~kHz (500~Hz), the minimum NEP is 81.0~pW/$\sqrt{\text{Hz}}$ (3.5~nW/$\sqrt{\text{Hz}}$) at a bias voltage of 1~V and 4.3~nW/$\sqrt{\text{Hz}}$ (44.1~nW/$\sqrt{\text{Hz}}$) at zero bias.\\

\begin{table*}[h]
\begin{tabularx}{\textwidth}{cP{42mm}rS[table-format=3.2]rccr}
\toprule
 & detector + antenna & $f$ [GHz] & $\gamma_{THz}$ \Big[$\frac{V}{W}$\Big] & NEP \Big[$\frac{\text{pW}}{\sqrt{\text{Hz}}}$\Big] & diode & flexible & ref.\\
\midrule
%\includegraphics[height=8pt]{pic/hex.eps} & GFET + log-periodic & 300 & 0.15 & 200000 &  &  & \cite{vicarelli12}\\
%\includegraphics[height=8pt]{pic/hex.eps} & GFET + log-periodic & 263-353 & 0.25 & 80000 &  &  & \cite{bianco15}\\
%\rowcolor[gray]{.9} \includegraphics[height=8pt]{pic/hex.eps} & rigid 1D-MIG + bowtie & 170 & 0.4 & 90000 & \checkmark &  & this work \\
%\includegraphics[height=8pt]{pic/hex.eps} & bilayer GFET & 290-380 & 1.2 & 2000 &  &  & \cite{spirito14}\\
 & MIM + bowtie$^*$ & 28300 & 0.2 &  & \checkmark &  & \cite{Jayaswal2018}\\
 & traveling wave MIM + bowtie$^*$ & 28300 & 0.3 &  & \checkmark &  & \cite{Pelz2016}\\
\includegraphics[height=8pt]{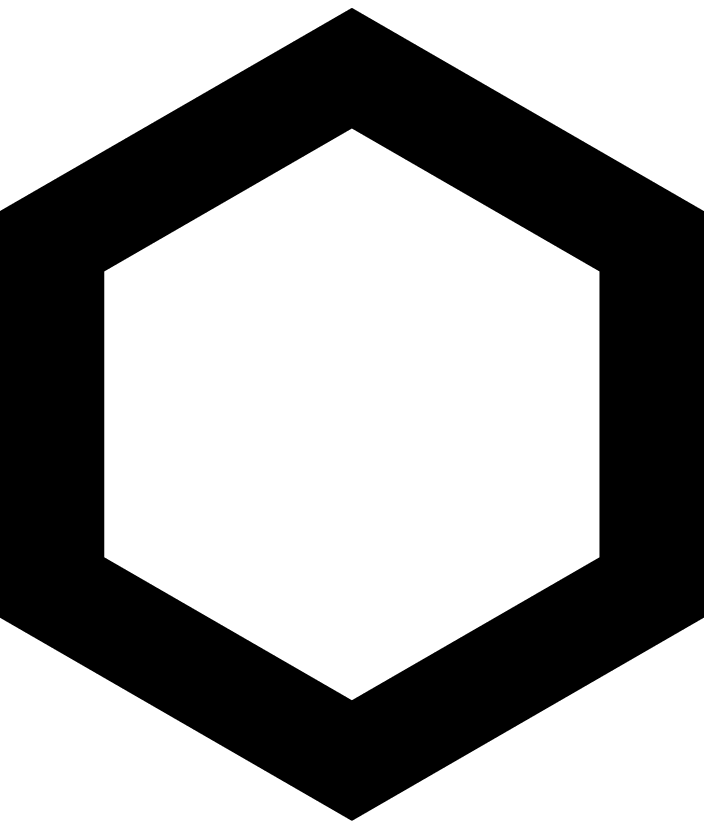} & GFET + bowtie & 487 & 2 & 3000 &  & \checkmark & \cite{yang17}\\
 & CNT Schottky & 540 & 2.6 & 20000 &  &  & \cite{Manohara2005}\\
\includegraphics[height=8pt]{pic/hex.eps} & photothermoelectric & 2519 & 10 & 1100 &  &  & \cite{cai14}\\
 & MSM + log spiral$^*$ & 300 & 10.8 & 100 & \checkmark &  & \cite{Moon2018}\\
\rowcolor[gray]{.9} \includegraphics[height=8pt]{pic/hex.eps} & flexible 1D-MIG + bowtie & 167 & 83.7 & 81 (3500) & \checkmark & \checkmark & \\
 & unipolar nanodiode + bowtie & 1500 & 300 & 330 & \checkmark & & \cite{balocco2011}\\
\includegraphics[height=8pt]{pic/hex.eps} & photothermoelectric$^*$ & 2519 & 715 & 16 &  &  & \cite{cai14}\\
\includegraphics[height=8pt]{pic/hex.eps} & ballistic diode + bowtie$^*$ & 685 & 764 & 34 & \checkmark &  & \cite{auton17}\\
% & Si nMOS FET + bowtie & 250-1050 & 5000 & 10 &  &  & \cite{schuster2011}\\
\bottomrule
\end{tabularx}
* responsivity calculated from incident or absorbed rather than emitted power
\caption{Comparison to other THz detectors from literature. The rectenna discussed in this work is highlighted in grey. The hexagon in the first column indicates if the detector is based on graphene. The detectors marked with an asterisk use the actual incident power on the device, rather than the emitted power of the source. The values reported for our detector come from the total emitted power $P_{ava}$ from the source and are thus likely underestimated. The NEP of the device from this work was measured at 250~kHz. The NEP at a frequency of 500~Hz is provided in parenthesis as a comparison.}
\label{table:comp}
\end{table*}

Table \ref{table:comp} compares our rectenna on a flexible substrate to other THz detectors from literature.
The rectenna shows a higher responsivity than MIM or metal-semiconductor-metal (MSM) based rectennas, despite being fabricated on a flexible substrate.
It also outperforms the only other reported flexible THz detector by Yang et al. based on a graphene field-effect transistor \cite{yang17} and most other THz detectors based on the absorption effects.
It should be noted that many references derive values for the optical responsivity from the incident or absorbed power on the device area.
Since our reported optical responsivity uses the \textit{total emitted} power $P_{ava}$ from the source (not scaled to the device area), it represents a lower bound that is mainly limited by the impedance mismatch between diode and antenna and absorption and scattering in the beam path.\\

\section*{}
%Conclusion

In summary, we have demonstrated detection of THz signals through an edge-contacted graphene-based rectenna in free space measurements.
The flexible rectenna, fabricated using scalable, thin-film compatible processes and CVD-grown graphene, reaches a responsivity of 83.7~V/W at 167~GHz with a NEP of 81.0~pW/$\sqrt{\text{Hz}}$, enabling low-cost energy harvesting at the chip level in flexible electronics.\\

\section{Author Information}
\subsection{ORCID}

\begin{itemize}
\item Andreas Hemmetter: 0000-0002-0932-2183
\item Xinxin Yang: 0000-0003-4464-6922
\item Zhenxing Wang: 0000-0002-2103-7692
\item Martin Otto: 0000-0001-6704-3468
\item Burkay Uzlu: 0000-0001-6776-8901
\item Marcel Andree: 0000-0003-3645-0831
\item Ullrich Pfeiffer: 0000-0002-6753-7879
\item Andrey Vorobiev: 0000-0003-2882-3191
\item Jan Stake: 0000-0002-8204-7894
\item Daniel Neumaier: 0000-0002-7394-9159
\item Max C. Lemme: 0000-0003-4552-2411
\end{itemize}

\subsection{Author contributions}
The experiments have been conceived by A.H., Z.W. ,M.L. and D.N.
A. H. designed the devices, and performed the electrical characterization. 
A.H., M.O. and B.U. fabricated the devices.
X. Y. performed the electrical and optical characterization. M. A. performed the noise measurement.
All authors contributed to discussions and the analysis and interpretation of the results. All authors have given approval to the final version of the manuscript.

\subsection{Notes}
The authors declare no competing financial interest.

\begin{acknowledgement}
This research was financially supported by the EU projects Graphene Flagship (Contract No. 881603) and WiPLASH (Contract No. 863337) and the DFG projects HiPeDi (Contract No. WA4139/1.1) and GLECS-2 (Contract No. NE1633/3).
\end{acknowledgement}

\bibliography{main}

\end{document}